\documentclass[els,twocolumn,amsmath,amssymb,keywords,floatfix]{revtex4}
\usepackage[ansinew]{inputenc}
\usepackage{multirow}
\usepackage{mathrsfs}
\usepackage{revsymb}
\usepackage{slashed}
\usepackage{longtable}
\usepackage[T2A]{fontenc}
\usepackage{amsmath}
\usepackage{amssymb}
\usepackage{bm}
\usepackage{epsfig}
\usepackage{graphicx}
\usepackage{color}

\begin{document}

\title{State-Selective High-Energy Excitation of Nuclei by Resonant Positron Annihilation}

\author{N.~A.~Belov}
\affiliation{Max Planck Institute for Nuclear Physics, 69117 Heidelberg, Germany}
\author{Z.~Harman}
\affiliation{Max Planck Institute for Nuclear Physics, 69117 Heidelberg, Germany}

\begin{abstract}
In the annihilation of a positron with a bound atomic electron, the virtual $\gamma$ photon created may
excite the atomic nucleus. We put forward this effect as a spectroscopic tool for an energy-selective
excitation of nuclear transitions. This scheme can efficiently populate nuclear levels of arbitrary
multipolarities in the MeV regime, including giant resonances and monopole transitions.
In certain cases, it may have a higher cross sections than the conventionally used Coulomb excitation
and it can even occur with high probability when the latter is energetically forbidden.

\begin{keywords} Positron annihilation, Nuclear excitation, Giant Resonances \end{keywords}
\end{abstract}

\date{\today}



\maketitle

\begin{figure}[t!]
\begin{center}
\includegraphics[width=0.45 \textwidth,angle=0]{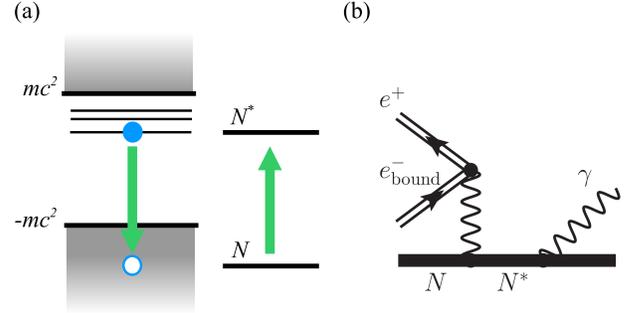}
\end{center}
\caption{\label{fig:VD1} (Color online.) (a) The illustration of positron annihilation via
nuclear excitation by the corresponding fermionic and nuclear level schemes. $mc^2$ is the rest energy of the $e^-$, and $N$ ($N^*$) stands for the nuclear
initial (excited) state.
(b) The lowest-order Feynman diagram of the NERPA process followed by nuclear de-excitation by emission of a photon. Thick lines denote
nuclear states, double lines denote fermions in the Coulomb field of the nucleus, and wave lines represent
real or virtual photons.
}
\end{figure}

Positron collisions with atomic matter lead to a number of processes~\cite{And10,Adr13,Ade11,Mue06,Sur05},
among which annihilation with shell electrons is one of the most prominent effects.
Typically, annihilation leads to the emission of $\gamma$ rays.
Alternatively, the atomic nucleus may resonantly absorb the whole energy of the annihilating particles and become excited,
as first proposed in~\cite{Pre52}. 
We refer to this single-step process as nuclear excitation by resonant positron annihilation 
(NERPA)~\cite{NEPEA}.
It is represented in Fig.~\ref{fig:VD1}(a) by the level schemes of the
electron shell and the nucleus. NERPA may be followed by radiative nuclear de-excitation.
The leading diagram of this two-step process, labeled as NERPA-$\gamma$, is shown in Fig.~\ref{fig:VD1}(b).

Attempts to observe NERPA have not been conclusive so far.
Only an upper bound of its cross section has been determined in the latest experiment~\cite{Cas01}.
In this measurement, a monochromatic positron beam was employed, and the results suggest that in previous
experiments employing broadband sources~\cite{Sai94}, some unidentified non-resonant process may have played the dominant role,
and, therefore, yielded cross sections well above the theoretical NERPA cross sections~\cite{Pre52,Gre78}.
Further works~\cite{Pis85,Hor88}, as well as newer results~\cite{Kal10} have improved
the theoretical description. However, the disagreement of different measurements is still not understood.
Therefore, our studies and future experimental work is anticipated to provide an unambiguous identification of the NERPA process
and a determination of its cross section.

NERPA constitutes a way to excite nuclei which is alternative to photo- and Coulomb excitation.
Photo-excitation experiments are conventionally done with bremsstrahlung, synchrotron or inverse Compton sources.
X-ray free electron lasers, providing the highest photon intensities, are presently limited to the keV photon energy regime~\cite{LCLS,XFEL}.
The great advantage of photo-excitation is the monochromaticity of the x- or $\gamma$-ray beam and the resonant
character of the nuclear excitation. The transitions predominantly accessible this way are, however, of electric-dipole ($E1$) type.
On the other hand, Coulomb excitation, i.e. excitation by inelastic scattering of massive charged particles
may induce transitions of arbitrary multipolarities. As an example, octupole deformations have been recently investigated
by Coulomb excitation in collisions with lighter nuclei, and monitoring the subsequent $\gamma$ decay~\cite{Gaf13}.
Coulomb excitation is however not selective with respect to the nuclear energy levels.

In this Letter we show that NERPA has an attractive combination of the above advantages:
the resonant character of the excitation and a significant cross section regardless of the multipolarity.
We find that in certain cases its cross section can be comparable to or an order of magnitude higher than that of Coulomb 
excitation, leading to several applications. One of examples is the possibility to resonantly induce monopole transitions,
that is particularly important for the studies of deformed nuclei.
Another application is a collective nuclear excitation, the giant monopole resonance~\cite{Spe81,Wam77},
also termed as ``breathing" mode as it involves an oscillation of the nuclear volume, and provides
the only way known for the experimental study of nuclear compressibility.
In the case of giant nuclear resonances of any multipolarity, NERPA bears all the above-mentioned advantages of Coulomb excitation.
A further property of NERPA is that experiments may be performed with neutral atoms, e.g. solid targets,
without the necessity of stripping off the atomic electrons or accelerating the nuclei.

Very recently, intense positron jets with MeV energies or above have been generated in laser-plasma interactions~\cite{Che10,Mul09,Sar13,Sar14}, with
a quasi-monoenergetic spectrum, and with positron numbers reaching $10^{10}$ per shot~\cite{Che10}.
In this Letter we therefore put forward the indirect laser excitation of nuclei via NERPA, utilizing positrons produced by strong laser pulses.
This scheme is complementary to excitation by $\gamma$ photons generated via Compton backscattering as planned, e.g., for the
ELI facility~\cite{ELI,Mou11}, and to other direct or indirect nuclear excitation mechanisms utilizing optical \cite{Led03,McK03,Spo08,Koc99,DiP12}
or x-ray light sources~\cite{Vag14,Roh12,Gun14,Hee13,Bur06}.

The NERPA process is the time reverse of $e^- e^+$ internal pair conversion accompanying the decay of an excited nucleus.
A concise model of that process is given in~\cite{SSG81_1,SSG81_2}. We extend the formalism of these works for the case
of NERPA. Its transition rate (probability per unit time) is given by
\begin{eqnarray}
P_{\rm N}^{j',j}&=&\frac{\delta(E+E'-\omega)}{[J_i,j',j]}
\sum_{M_i=-J_i}^{J_i}\sum_{M_f=-J_f^*}^{J_f^*}
\sum_{\mu=-j}^{j}\sum_{\mu'=-j'}^{j'} \\
&\times&(2\pi)^2\left|\alpha \int d {\bf r}_n \int d {\bf r}_f j_n({\bf r}_n) j_f({\bf r}_f)
\frac{e^{i\omega|{\bf r}_n-{\bf r}_f|}}{|{\bf r}_n-{\bf r}_f|}\right|\,, \nonumber
\end{eqnarray}
where $j$ and $j'$ are the total angular momentum quantum numbers of the $e^+$ and $e^-$ states, respectively,
$J_i$ and $J_f^*$ are angular momenta of the nuclear initial and final states, and the $M$-s and $\mu$-s are the associated magnetic quantum numbers.
Furthermore, $\alpha$ is the fine-structure constant, ${\bf r}_n$ and ${\bf r}_f$ denote nuclear and fermionic coordinates, and
$\omega$ stands for the virtual $\gamma$ energy. $E$ and $E'$ are the $e^+$ and $e^-$ energies, respectively, and $j_n$ and $j_f$ are nuclear
and fermionic 4-currents~\cite{LL_IV}. Also, the notation $[j_1,...,j_n]=\prod_{i=1}^{n}(2j_i+1)$ has been introduced.

The further derivation involves a splitting of the double integral over nuclear and fermionic coordinates
into a product of nuclear and fermionic parts.
This can be performed in the non-penetration approximation, where
the overlap integral of the bound-electron wave function within the nuclear volume is neglected.
This simplification is generally used in the calculation of quantities related to the interaction of shell
electrons with the nucleus \cite{SSG81_1,SSG81_2,Pal07a,Pal08}.
Then it is possible to factorize the NERPA transition rate as
$P_{\rm N}^{j',j}=\beta_{\rm N}^{j',j}P_{\gamma}$,
where $P_{\gamma}$ is the rate for the state reached by NERPA
to decay by $\gamma$ emission into the initial state.

For the coefficient $\beta_{\rm N}^{j',j}$ one may obtain expressions for any electric ($\lambda=E$) or
magnetic ($\lambda=M$) nuclear transition multipolarity,
i.e. for any (non-zero) values of the angular momentum $L$ of the virtual photon:
\begin{equation}
\beta_{\rm N}^{j',j}(\lambda L)=\frac{[J_f]}{[J_i,j',j]}\sum_{\kappa\kappa'}\frac{4\pi\alpha\omega}{L(L+1)} s |\kappa\kappa'|
\rho_\lambda,
\end{equation}
where the radial part for the $\lambda=E$ case is $\rho_E=|(\kappa-\kappa')(R_3+R_4)+L(R_1+R_2+R_3-R_4)|^{2}$,
and for the $\lambda=M$ case $\rho_M=|(\kappa+\kappa')(R_3+R_4)|^{2}$.
We have also introduced the following notation in terms of a 3-$j$ symbol~\cite{Edmonds}:
$s=\left(%
\begin{array}{ccc}
  j & j' & L \\
  \frac{1}{2} & -\frac{1}{2} & 0 \\
\end{array}%
\right)^{2}$.
For the Dirac angular momentum quantum number $\kappa$ ($\kappa'$) of the positron (electron) holds:
$
||\kappa|-|\kappa'||\leq L\leq |\kappa|+|\kappa'|-1$,
$j=|\kappa|-1/2$, $j'=|\kappa'|-1/2$.
The radial integrals $R_a$, $a \in \{1,\dots,6\}$ are defined as in~\cite{SSG81_1} with the analytical
form of the radial Coulomb-Dirac wave functions for bound and free particles.
All results are obtained here for the $1s_{1/2}$ electron orbital, yielding the highest rate of NERPA.

\begin{figure}[t!]
\includegraphics[width=0.35 \textwidth]{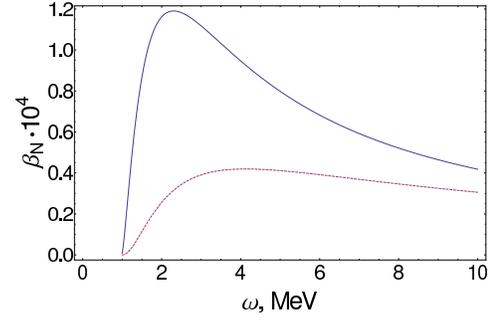}
\caption{\label{fig:NERPA} (Color online.) The NERPA coefficient $\beta_{\rm N}^{j',j}$ for $E2$ nuclear transition between
$1/2+$ and $5/2+$ nuclear states
vs. the photon energy $\omega$ for $Z=50$ ($^{115}$Sn). Blue solid line: the case of $j=L-1/2$, brown dashed line: $j=L+1/2$.}
\end{figure}

Fig. \ref{fig:NERPA} shows the calculated dependence of the NERPA coefficient $\beta_{\rm N}^{j',j}$
for an electric-quadrupole ($E2$) nuclear transition between $1/2+$ and $5/2+$ states on the virtual photon energy $\omega$ in $^{115}$Sn.
Curves are given for two different possible positron angular momenta $j=L\pm1/2$;
both values $\beta_{\rm N}^{j',L\pm1/2}$ have to be taken into account in the calculation of the cross section.
One can see that nuclear transitions about 2~MeV, corresponding to the maximum
of these curves, are preferred. This behaviour is analogous to the energy dependence of the bound-free pair creation
cross section in heavy ion collisions~\cite{Bec87}.

\begin{figure}[t!]
\includegraphics[width=0.45 \textwidth]{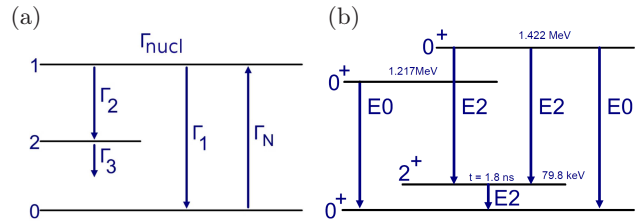}
\begin{picture}(0,0)(100,100)
\put(-140,175){(a)}\put(-20,175){(b)}
\end{picture}
\caption{\label{fig:NERPA_E0} (Color online.) (a)
Nuclear level scheme for the observation of NERPA. Level 0 denotes the initial state, level 1 is the state populated by
NERPA, and level 2 is the final metastable state. The resonance widths are assumed to satisfy $\Gamma_3\ll \{\Gamma_1,\Gamma_2\}$.
(b) The lowest $0^+$ and $2^+$ nuclear levels in $^{168}_{68}$Er. See text for details.
}
\end{figure}

A promising approach to observe the NERPA process is to measure the photon emitted by the decaying nucleus,
but it is a comparably weak effect superimposed on the background of other photon emission processes in the studied system.
To circumvent this problem, as in Refs.~\cite{Pre52,Gre78},
 we consider nuclei with a long-living state, which can be populated by
some radiative transition from the state excited by NERPA. The decay of this metastable state is the
signature of the NERPA process, which may be measured with some time delay after the excitation takes place.
The level scheme of the nucleus may be the one presented on Fig.~\ref{fig:NERPA_E0}~(a).
Within this scheme, the state labeled as $2$ is metastable, meaning that $\Gamma_3\ll \{\Gamma_1,\Gamma_2\}$, where the $\Gamma$-s
denote transition widths as in Fig.~\ref{fig:NERPA_E0}(a). $\Gamma_{\rm N}$ stands for
the resonance width of the NERPA transition, which is related to the $\gamma$ emission width as
$\Gamma_{\rm N}=\hbar P_{\rm N}=\sum_j[j]\beta_{\rm N}^{1/2,j}\Gamma_1$. The cross section of the total process
in dependence on the positron energy $E$ (and corresponding momentum $p$)
is given by the expression (for a derivation of cross sections of nuclear-resonant
processes, see e.g.~\cite{Pal07a,Pal08}):
\begin{equation}
\begin{split}
\label{eqn:CrossSection}
\sigma_{\rm N \gamma}(E)=\frac{2\pi^2 \hbar^2}{p^2} \frac{\Gamma_2^\gamma}{\Gamma_{\rm nucl}}
\frac{\Gamma_{\rm tot}/(2\pi)}
{(E-E_{\rm res})^2+\Gamma_{\rm tot}^2/4} \Gamma_{\rm N}\,,
\end{split}
\end{equation}
where $E_{\rm res}$ is the resonance energy, $\Gamma_2^\gamma$ is the width
of the $\gamma$ decay of the final NERPA state to the metastable state. This
width is connected to the internal conversion (IC) coefficient $\alpha_{\rm IC}$ through $\Gamma_2^\gamma=\Gamma_2/(1+\alpha_{\rm IC})$.
$\Gamma_{\rm nucl}=\hbar/\tau$ is the total width of the considered nuclear level, expressed in terms of its lifetime $\tau$. The total resonance
line width is given as the sum of the nuclear and atomic $K$-shell widths~\cite{Kol90}: $\Gamma_{\rm tot}=\Gamma_{\rm nucl}+\Gamma_K$. The excitation
cross section $\sigma_{\rm N}$ can be obtained from the above formula by substituting the branching ratio ${\Gamma_2^\gamma}/{\Gamma_{\rm nucl}}$ by~1.

\begin{table}[t!]
\caption{\label{tab:NERPA} Data for different elements which suit the nuclear level scheme on Fig.~\ref{fig:NERPA_E0}~(a).
Notations are as defined there and in the text. The symmetry (``symm.") of the nuclear levels involved -- total angular momentum, parity --
is given. Nuclear data are taken from Ref.~\cite{NDS}.}
\begin{tabular}{cccc}
  \hline
  \hline
  Element & $^{174}_{72}$Hf  & $^{115}_{50}$Sn  & $^{115}_{49}$In\\
  \hline
  $E_{\rm res}$ (keV)                   & 281 & 430 & 90\\
  level 0 symm. & $0^+$ & $1/2^+$ & $9/2^+$ \\
  level 1 (keV); symm. & 1227; $2^+$ & 1417; $5/2^+$& 1078; $5/2^+$ \\
  level 2 (keV); symm. & 91; $2^+$ & 613; $7/2^+$ & 336; $1/2^+$\\
  lifetime of level $2$, $\tau_2$
                      & 1.66 ns & 3.26 $\mu$s & 4.5 h\\
  $\Gamma_1$ (eV)   & $9.00\cdot10^{-4}$ & $1.39\cdot10^{-3}$ & $5.54\cdot10^{-4}$ \\
  $\Gamma_2^\gamma$ (eV)     & $4.04\cdot10^{-4}$ & $9.9\cdot10^{-5}$ & $1.05\cdot10^{-4}$\\
  $\Gamma_3$ (eV)  & $3.965\cdot10^{-7}$ & $2.02\cdot10^{-10}$ & $4.06\cdot 10^{-20}$\\
  $\beta_{\rm N}$  & $7.6\cdot10^{-5}$ & $4.1\cdot10^{-5}$ & $1.9\cdot10^{-6}$\\
  $\Gamma_{\rm nucl}$ (eV)  & $1.83\cdot10^{-3}$ & $1.88\cdot10^{-3}$ & $6.6\cdot10^{-4}$\\
  $\Gamma_{K}$ (eV) & 33.0 & 7.9 & 7.3 \\
  $\sigma_{\rm N}(E_{\rm res}) $ (b) & $1.1\cdot10^{-4}$ & $1.5\cdot10^{-4}$ & $1.7\cdot10^{-5}$ \\
  $\sigma_{{\rm N}\gamma}(E_{\rm res})$ (b) & $6.1\cdot10^{-6}$ & $3.1\cdot10^{-6}$ & $1.1\cdot10^{-6}$ \\
  \hline
  \hline
\end{tabular}
\end{table}

In Tab.~\ref{tab:NERPA} results for 3 different elements are presented, namely,
$^{174}$Hf, $^{115}$Sn, and $^{115}$In. All these elements possess the level scheme presented
on Fig.~\ref{fig:NERPA_E0}~(a), and an $E2$ nuclear transition for the NERPA process.
For $^{174}$Hf, the cross section $\sigma_{{\rm N\gamma}}$ is the largest. However, here, the metastable
level has a comparably short lifetime $\tau_2$, which may not be sufficient for an experimental
differentiation of the signal from the direct bound-free $e^- e^+$ photoannihilation process.
The cross section and $\tau_2$ for the case of $^{115}$Sn and $^{115}$In are most suitable
for a possible measurement. The NERPA process is significantly enhanced for some elements such as e.g. $^{174}$Hf and $^{115}$Sn.
Besides the increase of the $\beta$ coefficient with $Z$, it is also because for these isotopes, the nuclear transition energy renders $\beta$
nearly maximal (see Fig.~\ref{fig:NERPA}). In the case of $^{115}$In, it is rather the low positron resonance kinetic energy that boosts the
total NERPA cross section due to the pre-factor $1/p^2$ in Eq. (\ref{eqn:CrossSection}).

\begin{table*}[t!]
\caption{\label{tab:CNO} Results for a range elements. $\Delta E$ denotes the energy of the excited nuclear state,
$\Gamma_{\gamma}$ and $\Gamma_{\rm nucl}$ are its radiative and total decay width, respectively.
See the previous table for notations and the text for further details.}
\begin{tabular}{ccccccccccc}
  \hline
  \hline
  Element & $^{12}_{6}$C & $^{14}_{7}$N  & $^{16}_{8}$O & $^{56}_{26}$Fe & $^{38}_{19}$K & $^{14}_{7}$N & $^{18}_{9}$F & $^{18}_{9}$F & $^{21}_{10}$Ne & $^{208}_{82}$Pb \\
  \hline
  $\Delta E$ (keV)      & 4439 & 2313 & 6917 & 2958 & 2646 & 5691 & 1081 & 2101 & 2788 & 13500 \\
  $E_{\rm res}$ (keV)   & 3418 & 1289 & 5899 & 1947 & 1629 & 4670 &   60 & 1080 & 1767 & 12600 \\
  Multipolarities   & $2^+\overset{E2}{\rightarrow} 0^+$ & $1^+\overset{M1}{\rightarrow} 0^+$ & $0^+\overset{E2}{\rightarrow} 2^+$ & $0^+\overset{E2}{\rightarrow} 2^+$
     & $3^+\overset{E1}{\rightarrow} 4^-$ & $1^+\overset{E1}{\rightarrow} 1^-$
     & $1^+\overset{E1}{\rightarrow} 0^-$ & $1^+\overset{E1}{\rightarrow} 2^-$ & $3/2^+\overset{E1}{\rightarrow} 1/2^-$ & $0^+\overset{E1}{\rightarrow} 1^-$ \\
  $\beta_{\rm N}$ & $2.1\cdot 10^{-8}$ & $9.1\cdot 10^{-10}$ & $9.1\cdot 10^{-7}$ & $3.2\cdot 10^{-5}$ & $3.2\cdot 10^{-6}$ & $1.2\cdot 10^{-7}$ 
     & $2.1\cdot 10^{-8}$ & $5.3\cdot 10^{-7}$ & $2.2\cdot 10^{-7}$ & $9.6\cdot 10^{-5}$ \\
  $\Gamma_{\rm nucl}$ (eV) & $1.08\cdot 10^{-2}$ & $9.7\cdot 10^{-3}$ & $1.4\cdot 10^{-1}$ &$2.4\cdot 10^{-2}$ & $6.7\cdot 10^{-7}$ & $6.0\cdot 10^{-2}$ 
     & $3.4\cdot 10^{-5}$ & $1.9\cdot 10^{-4}$ & $8.1\cdot 10^{-6}$ & $4\cdot 10^{6}$ \\
  $\Gamma_\gamma$ (eV) & $1.08\cdot 10^{-2}$ & $9.7\cdot 10^{-3}$ & $1.4\cdot 10^{-1}$ &$5.1\cdot 10^{-4}$ & $6.6\cdot 10^{-7}$ & $2.2\cdot 10^{-2}$ 
     & $3.4\cdot 10^{-5}$ & $7.1\cdot 10^{-5}$ & $1.4\cdot 10^{-6}$ & $4\cdot 10^{6}$\\
  $\Gamma_{K}$ (V) & $2\cdot 10^{-3}$ & $3.6\cdot 10^{-3}$ & $6\cdot 10^{-3}$ & $6.1\cdot 10^{-1}$ & $1.8\cdot 10^{-2}$ & $3.6\cdot 10^{-3}$
    & $9.5\cdot 10^{-3}$ & $9.5\cdot 10^{-3}$ & $1.4\cdot 10^{-2}$ & $5.5\cdot 10^{1}$\\
  $\sigma_{\rm N}$  (b) & $1.1\cdot 10^{-5}$ & $2.2\cdot 10^{-5}$ & $2.1\cdot 10^{-4}$ & $4.4\cdot 10^{-5}$ & $2.6\cdot 10^{-7}$ & $1.5\cdot 10^{-5}$
     & $1.2\cdot 10^{-5}$ & $1.7\cdot 10^{-5}$ & $4.1\cdot 10^{-8}$ & $5.5\cdot 10^{-3}$\\ 
  $\sigma_{{\rm C}}$ (b) & 6.7 & 0.039 & 9.4 & 2.1 & $2.4\cdot 10^{-8}$ & $7.1\cdot 10^{-5}$ 
     & $1.9\cdot 10^{-5}$ & $5.3\cdot 10^{-6}$ & $4.2\cdot 10^{-8}$ & $6.29\cdot 10^{2}$ \\
  \hline
  \hline
\end{tabular}
\end{table*}

We separately discuss the case of electric monopole (E0) nuclear transitions excited by the NERPA process.
These $0^+\rightarrow 0^+$ transitions are important for the investigation of the spectrum of deformed nuclei, since
$0^+$ states in heavy deformed nuclei~\cite{Wal01} are formed by a superposition of many oscillation modes and are objects of interest for
microscopic nuclear models~\cite{Buc07}.
The direct one-photon nuclear excitation is forbidden in this case, therefore we
compare the first (nuclear) step of the NERPA
process to the dominant allowed nuclear decay channel of $0^+\rightarrow 0^+$ nuclear transitions, namely, IC.
The ratio $\delta_{\rm N}^{j',j}$ of the probabilities of the NERPA process and IC in the non-penetration approximation can be written as
$P_{\rm N}^{j',j}=\delta_{\rm N}^{j',j}P_{\rm IC}$. Transition rates for the E0 case can be derived with the help of, e.g., Ref.~\cite{SSG81_2}.

E0 NERPA excitations enable the investigation of $0^+$ level cascades in heavy deformed nuclei.
$^{168}$Er is a typical example of such nuclei, having a $0^+$ ground state and a $0^+$ excited state at an energy of
1.217~MeV, which can decay by $E2$ $\gamma$ emission to a metastable $2^+$ level at an energy of 79.8~keV with a lifetime of 1.853~ns
(see Fig.~\ref{fig:NERPA_E0}~(b)).
Here, the $\gamma$-decay of the metastable level follows the NERPA process.
For the described level scheme in~$^{168}$Er, the lifetime of the excited $0^+$ level and the intensity of $0^+\rightarrow 0^+$
transition are unknown, however, we can calculate the rate of IC transition between these levels, for instance following Ref.~\cite{Chu56}, and then,
with the help of the $\delta$ coefficient derive the rate and width of the NERPA transition.
Thus one can obtain the resonance cross section of the total NERPA-$\gamma$ process through the reduced transition probability $B_{E2}$~\cite{BM98}
of the $0^+\rightarrow 2^+$ de-excitation, yielding
$\sigma_{\rm N\gamma}(E_{\rm res})~[{\rm b}]={1.12 \cdot 10^{-11}}/({B_{E2}}~[e^2{\rm fm}^{4}])$.
The cross section may be determined experimentally, allowing one to obtain the reduced $E2$ probability for the second step
of the NERPA process. This quantity is connected to the nuclear deformation parameter, thus the latter can be inferred
from NERPA measurements and may be used to benchmark nuclear structure models.

Besides offering an alternative technique of nuclear spectroscopy, the theoretical and experimental
investigation of NERPA is of relevance for applications connected with positron-matter interaction.
Firstly, the NERPA process is potentially relevant for cosmic ray studies
with positron interacting with the atoms in atmosphere (see e.g.~\cite{Adr13}).
Furthermore, in experimental investigations of nuclear reactors,
NERPA may influence the dynamics of chain reactions involving $\beta^+$ emitters.
A similar situation may occur in star evolution simulations \cite{Ade11}.
In both latter cases, the positron interacts with heavy atoms or ions, in which
the NERPA strength is boosted.

Light and medium-$Z$ elements such as C, N, O and Fe may also feature significant NERPA cross sections. The light elements typically
occur in biological environments and in solar plasmas. In addition, Fe is present in large quantities in the radiation zone of the Sun.
The light atoms are also of importance in cosmic ray studies and in medical positron tomography research.
Table~\ref{tab:CNO} shows widths and resonance cross sections for the one-step NERPA process for these elements $\sigma_{\rm N}$.
For O or  Fe, the NERPA cross section $\sigma_{\rm N}$ is even larger than in the case of $^{115}$In, the element which was investigated
earlier (see Tab.~\ref{tab:NERPA}).
For comparison, cross sections for Coulomb excitation $\sigma_C$ with high-energy protons (or positrons) are given.
NERPA is in several cases comparable to or up to an order of magnitude stronger than Coulomb excitation.

We  have identified the strongest NERPA excitation so far for a 13.5-MeV giant dipole resonance (GDR) in~$^{208}$Pb.
Since for such resonances the nuclear level width is in the MeV range, these resonances also feature the highest integrated NERPA
cross sections.
For instance, the estimation of the NERPA cross section integrated with the effective resonance width
reaches  $\int \sigma_{\rm N}(E) dE=3.4 \cdot 10^4$~b$\cdot$eV for the mentioned GDR in $^{208}$Pb,
exceeding previous values for e.g. $^{115}$In by 8 orders of magnitude.
GDRs can be efficiently excited even with a broadband positron source generated
in laser-plasma interactions by existing novel methods~\cite{Sar13,Sar14}. E.g., for a positron beam with a 0.5-MeV width,
Coulomb excitation of the GDR is not significant, since it requires positron kinetic energies higher by $2mc^2$ than NERPA.
Therefore, exciting GDRs with laser-generated positron beams may be a viable alternative to observe NERPA,
circumventing the difficulties caused by the low nuclear line width in elements such as $^{115}$In~\cite{Cas01}.
Furthermore, NERPA can be used for the excitation of a certain energy region of a giant resonance,
enabling the investigation of the thermal evolution of the GDR width and structure.
Normally such studies are done by Coulomb excitation~\cite{End92,Kel99} which does not allow a selective excitation
but an energy-selective detection of subsequent $\gamma$ decays.
The problem of the theoretical description of the excitation of the certain part of the GDR is caused
by the complicated resonance structure consisting of the superposition of the several nuclear states.
To calculate the integrated excitation cross section in this case, one has to use some model
for the GDR strength function. For instance, it can be the double-Lorentzian model of the strength function~\cite{Bec92, Hel14}
for the case of the wide incoming positron distribution; or some more precise model, e.g.~\cite{Avd11}.

Another application of giant resonances where the energy selectivity of NERPA (due to its resonance structure) is beneficial
is the feeding of highly deformed states by the low-energy GDR component~\cite{Kmi05}.
This feeding effect is relevant for the investigation of the low-energy limit of the radiative dipole strength~\cite{Lit13},
which, in turn, has consequences in astrophysical simulations of the r-process nucleosynthesis of exotic nuclei~\cite{Gor04}.

In summary, positron annihilation with a shell electron is put forward as an alternative means to induce nuclear transitions at the MeV level.
Laser-produced positron beams may provide sufficient flux for efficient nuclear excitation. Cross sections are typically
the largest in medium-$Z$ and heavy elements with the nuclear transition energy near the maximum of the cross section about 2~MeV.
This process is potentially relevant in numerous applications connected with positron-matter interaction, and
is also anticipated to provide one with a novel means for the investigation of the structure of deformed nuclei and
the energy spectrum of the giant resonances, as it can energy-selectively excite transitions of any multipolarity.

We acknowledge insightful conversations with Stanislav Tashenov and Christoph H. Keitel.

\bibliography{MY_PHD_2}

\end{document}